# Current-induced in-plane magnetization switching in biaxial ferrimagnetic insulator

Yongjian Zhou[1,4], Chenyang Guo[2,4], Caihua Wan[2], Xianzhe Chen[1], Xiaofeng Zhou[1], Ruiqi Zhang[1], Youdi Gu[1], Ruyi Chen[1], Huaqiang Wu[3], Xiufeng Han[2], Feng Pan[1] and Cheng Song[1*]

[1]Key Laboratory of Advanced Materials (MOE), School of Materials Science and Engineering, Tsinghua University, Beijing 100084, China

[2]Beijing National Laboratory for Condensed Matter Physics, Institute of Physics, University of Chinese Academy of Sciences, Chinese Academy of Sciences, Beijing 100190, China

[3]Institute of Microelectronics, Tsinghua University, Beijing 100084, China

[4]These authors contributed equally: Y. Zhou, C. Guo.
*songcheng@mail.tsinghua.edu.cn

**Ferrimagnetic insulators (FiMI) have been intensively used in microwave and magneto-optical devices as well as spin caloritronics, where their magnetization direction plays a fundamental role on the device performance. The magnetization is generally switched by applying external magnetic fields. Here we investigate current-induced spin-orbit torque (SOT) switching of the magnetization in $Y_3Fe_5O_{12}$ (YIG)/Pt bilayers with in-plane magnetic anisotropy, where the switching is detected by spin Hall magnetoresistance. Reversible switching is found at room temperature for a threshold current density of $10^7$ A $cm^{-2}$. The YIG sublattices with antiparallel and unequal magnetic moments are aligned parallel/antiparallel to the direction of current pulses, which is consistent to the Néel order switching in antiferromagnetic system. It is proposed that such a switching behavior may be triggered by the antidamping-torque acting on the two antiparallel sublattices of FiMI. Our finding not only broadens the magnetization switching by electrical means and promotes the understanding of magnetization switching, but also paves the way for all-electrically modulated microwave devices/spin caloritronics with low power consumption.**

## I. INTRODUCTION

Yttrium iron garnet ($Y_3Fe_5O_{12}$, YIG), a ferrimagnetic insulator (FiMI) with ultralow Gilbert damping and high permeability, has long term been applied in the microwave and magneto-optical devices. It has drawn increasing interest in spintronics field, such as the investigations of the spin Seebeck effect [1,2], spin Hall magnetoresistance (SMR) [3–5], spin pumping [4,6], non-local magnon transport [7], cavity magnon polariton [8] and magnon valves [9]. In particular, YIG is an ideal magnetic material for pure spin current transport because charge current and corresponding Seebeck effect and Nernst effect can be completely eliminated [2], providing a promising candidate for electronics with low energy dissipation. Note that the magnitude or even on/off state of signals and resultant device performance are strongly modulated by the magnetization direction in YIG-based applications, such as in spin Seebeck effect [2] and magnon valve [9]. From the application of spintronics, YIG is always treated as a simple "ferromagnet" with a single net moment, though it is a typical FiMI with two sublattices, especially a magnetic primitive cell of YIG contains 20 Fe moments and a complicated spin structure [10]. In such case, the external magnetic fields are generally used to switch the magnetization direction of YIG. A remarkable miniaturization trend on electronics calls for the switching of FiMI with a more convenient and efficient method.

The spin-orbit torque (SOT) in ferromagnet/heavy metal bilayers, where the angular momentum of spin current induced by charge current with spin Hall effect [11] is transferred into magnetic layer in the form of magnetic torque, provides an effective electrical means for manipulating magnetic dynamics and switching the uniform magnetization [12]. Previous studies concentrated on the SOT in metallic systems, including out-of-plane [13–16] and in-plane [17–19] magnetization switching, magnetic oscillations [20,21], domain wall motion [22,23] and skyrmions [24,25]. These concepts are transferred to the FiMI system, and there are several recent works reporting on the SOT switching in FiMI/heavy metal bilayers with perpendicular magnetic anisotropy [26–29], because of the lower energy dissipation and easy readout of the switching signal, such as by anomalous Hall effect. However, almost all of the microwave and spintronics applications of YIG, as mentioned above, are based on YIG with in-plane magnetization, therefore the electrical switching of in-plane magnetized YIG is strongly pursued (TABLE I). The experiments described here investigate the SOT switching of in-plane magnetized YIG (001) in YIG/Pt bilayers, where the two anti-parallel magnetic moments are set parallel/antiparallel to the direction of writing current.

**TABLE I. SOT switching in ferromagnet (FM), ferrimagnetic metal (FiM), ferrimagnetic insulator (FiMI), antiferromagnetic alloy (AFM) and antiferromagnetic insulator (AFMI) with out-of-plane and in-plane magnetic anisotropy.** Out-of-plane and in-plane SOT switching was extensively studied in FM and FiM. Out-of-plane switching of FiMI and in-plane switching of AFMI were realized recently. This work reports on the in-plane SOT switching of FiMI YIG.

| Magnetic anisotropy | FM | FiM | FiMI | AFM | AFMI |
|---|---|---|---|---|---|
| **Out-of-plane** | Refs. [13,14] | Refs. [15,16] | Refs. [26–29] | – | – |
| **In-plane** | Refs. [17,18] | Ref. [19] | This work | Refs. [30–34] | Refs. [35–37] |

## II. MRTHODS

The YIG films with in-plane magnetic anisotropy were deposited on GGG(001) substrates using a sputtering system with a base vacuum of $1 \times 10^{-6}$ Pa. After the deposition, high temperature annealing with oxygen atmosphere was carried out to further improve the crystalline quality and epitaxial relation between YIG film and GGG substrate [9]. The YIG thickness was determined using a pre-calibrated growth rate. The crystal structure was measured by x-ray diffraction (XRD). The in-plane magnetic anisotropy was recorded by vibrating sample magnetometer (VSM). The annealed YIG films were then transferred into another high-vacuum magnetron sputtering chamber to ex-situ deposit 5 nm Pt top layer at room temperature.

The YIG/Pt bilayers were patterned into eight-terminal devices with channel width of 5 μm through standard photolithography and Ar ion etching. The current induced magnetization switching measurements were carried out at room temperature by applying current pulses of $1.4 \times 10^7$ A cm$^{-2}$ with the width of 1 ms, then the transverse Hall resistance was recorded with a reading current of $1.2 \times 10^6$ A cm$^{-2}$. And the spin Hall magnetoresistance experiments with different current densities and magnetic fields were conducted with physical properties measurement system (PPMS).

**III. RESULTS AND DISSCUSION**

A series of YIG films ($t$ = 15, 20, 30, 60 nm) were grown on Gd$_3$Ga$_5$O$_{12}$ (GGG) (001) substrates by magnetron sputtering. In the following we focus primarily on data obtained from 20-nm-thick YIG films at room temperature. X-ray diffraction spectra in Fig. 1(a) shows that an additional peak from YIG (008) emerges in YIG/GGG sample, besides the diffraction peak from the GGG substrate, indicating that the YIG exhibit (001)-orientation, which serves as the basis of magnetization easy-plane (001). Figure 1(b) presents hysteresis loops of YIG with the magnetic field ($H$) applied along four in-plane directions of [100], [010], [110], and [$\bar{1}$10], as well as out-of-plane direction of [001]. A comparison of the squared in-plane loops and slanted out-of-plane loop shows that [001] is a hard-axis. The saturation field is ~15 Oe when $H$ is applied along [110] and [$\bar{1}$10], in contrast to ~75 Oe along [100] and [010]. This observation reflects that the YIG films possess fourfold in-plane magnetic anisotropy with easy-axes along [110] and [$\bar{1}$10] and hard-axes along [100] and [010], which results from the cubic anisotropy of bulk YIG [38]. The saturation magnetization ($M_S$) is around 115 emu cm$^{-3}$, which is lower than the bulk value (140 emu cm$^{-3}$) [39].

To perform current-induced in-plane magnetization switching measurements, the YIG were covered by 5-nm-thick Pt and then fabricated into eight-terminal devices with the channel width of 5 μm, where the writing current pulse channels are along easy-axes of [$\bar{1}$10] and [110] [Fig. 1(c)]. The in-plane switching measurements were carried out in the following way: five successive pulses (current density $J = 1.4 \times 10^7$ A cm$^{-2}$, 1-ms-width) were applied along [$\bar{1}$10] (write 1), and then along its orthogonal direction [110] (write 2) at zero external magnetic field. After each writing current pulse, a small reading current ($J = 1.2 \times 10^6$ A cm$^{-2}$) was applied and the transverse Hall resistance variation ($\Delta R_{xy}$) was recorded. $\Delta R_{xy}$ is intrinsically the spin Hall magnetoresistance of YIG/Pt system, where the spin polarization and relevant resistance in Pt can reflect the alignment of YIG moments [3]. Concomitant $\Delta R_{xy}$ for the magnetization switching is displayed in Fig. 1(e), where the red and blue circles correspond to the red (write 1) and blue (write 2) arrows, respectively (consistent correspondence for the following results). The current pulse of $1.4 \times 10^7$ A cm$^{-2}$ and 1-ms-width is the threshold current density for the switching [40]. The most eminent result is the two writing current pulses along [$\bar{1}$10] and [110] lead to the variation of $\Delta R_{xy}$ between low and high resistance states. Further inspection shows that $\Delta R_{xy}$ shows a sudden decrease (increase) and is almost saturated when the first write 1 (write 2) current pulse is applied, and the following four pulses cause a negligible variation in $\Delta R_{xy}$. This observation indicates step-like switching of in-plane magnetized YIG, which is most likely due to the small in-plane magnetic anisotropy of the present YIG films. Similar switching features were observed in antiferromagnetic $\alpha$-Fe$_2$O$_3$ [42] and Mn$_2$Au for the switching from hard- to easy-axis [34], but different from the multidomain switching in NiO/Pt [35]. The situation differs dramatically when the current pulses are applied along in-plane hard-axes ([100] and [010]) [Fig. 1(d)]. The $\Delta R_{xy}$ remains constant when the current pulses of $1.4 \times 10^7$ A cm$^{-2}$ are alternatively applied along [100] and [010] [Fig. 1(f)], indicating the negligible in-plane switching between the hard-axes. A similar behavior also occurs in YIG (111) films with in-plane isotropy [40]. It is then concluded that the magnetic anisotropy of YIG plays a fundamental role during in-plane switching process and the magnetic moments of in-plane biaxial YIG can be switched between the two easy-axes by a current pulse.

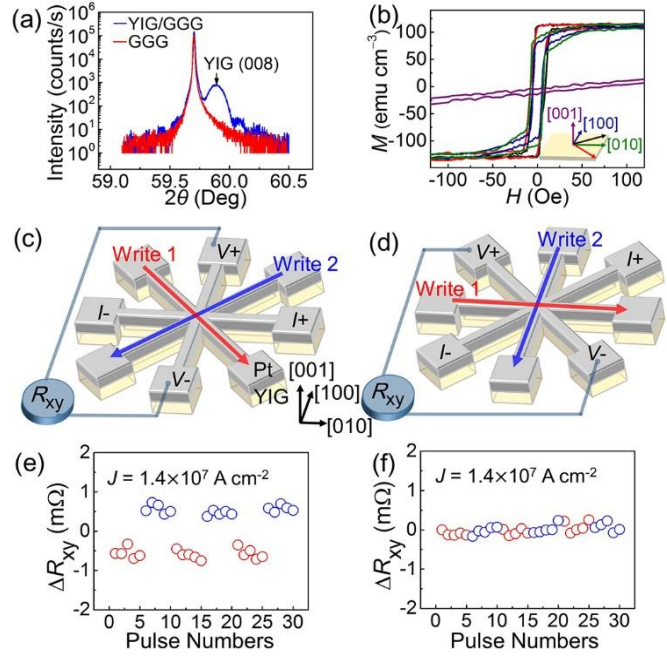

FIG. 1. Crystal structure, magnetism and current-induced magnetization switching of 20-nm-thick YIG. (a) X-ray diffraction spectra of GGG substrate and YIG/GGG sample. The peak from YIG (008) is marked. (b) Hysteresis loops at room temperature with magnetic field ($H$) applied along out-of-plane direction [001] and four in-plane directions, where [100] and [010] are in-plane hard-axes, while [110] and [$\bar{1}$10] are in-plane easy-axes. The axes are marked with the same color as their corresponding hysteresis loops. (c),(d) Measurement configurations of current-induced magnetization switching with writing current applied along in-plane easy- (c) and hard-axes (d), respectively. (e),(f) $\Delta R_{xy}$ as a function of the number of writing current pulses applied as depicted in (c) and (d), respectively.

The external magnetic field would provide an additional tool to modulate the current-induced switching of YIG. Figure 2(a) shows $\Delta R_{xy}$ as a function of current pulses with different additional fields ($H$ = 0, 5, 10, 50, and 75 Oe). The measurement configuration is identical to the one used in Fig. 1(c), except the $H$ is applied along hard-axis [100]. When $H$ is fixed at a quite low value of 5 and 10 Oe, which is just below and above the coercivity of YIG [Fig. 2(b)], there is no obvious difference as compared with the data at $H$ = 0. The scenario turns out to be different when $H$ increases to 50 Oe. The amplitude of $\Delta R_{xy}$ variation is reduced but still evident as $H$ is up to 50 Oe, which is close to the saturation field as shown in Fig. 2(b), suggesting that the current-induced in-plane magnetization switching is partially suppressed by $H$. Once $H$ increases to the saturation field of 75 Oe, there is negligible change of $\Delta R_{xy}$ when the current pulses alter their directions between write 1 and write 2, indicating that the current-induced magnetization switching is completely suppressed. This magnetic field modulated switching signals support that the present $\Delta R_{xy}$ variation is indeed ascribed to the SOT-induced in-plane switching of YIG ferrimagnetic sublattice magnetization, which is similar to the true Néel order switching in $\alpha$-$Fe_2O_3$ with magnetic field [42]. It indicates that thermal artifacts [42–44], which are at least not sensitive to a low magnetic field, have a negligible effect in our experiments. Moreover, both the unambiguous current-induced switching signals and artifacts are found in YIG/Cu/Pt trilayer [40].

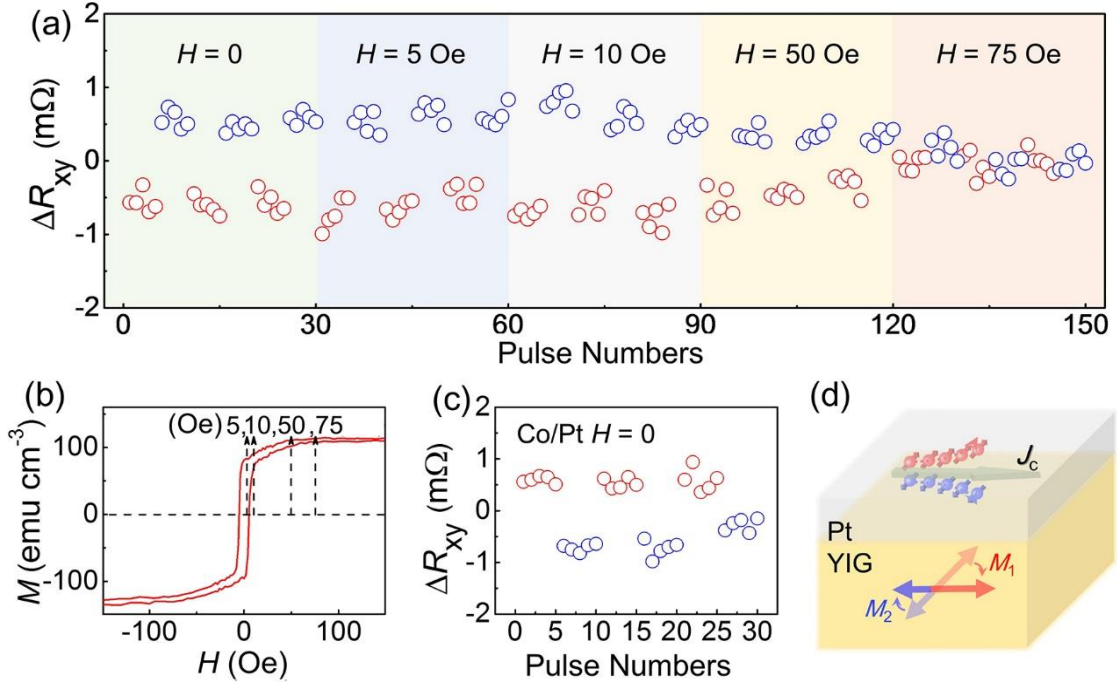

FIG. 2. SOT-induced switching in YIG/Pt and Co/Pt. (a) Summary of SOT-induced $\Delta R_{xy}$ as a function of writing current pulses with different additional fields $H$ applied along [100] in YIG/Pt bilayers. Results under different $H$ are separated by regions of different colors. (b) Hysteresis loop with $H$ applied along [100]. The typical $H$ employed in SOT switching measurements are denoted by dashed arrows. (c) $\Delta R_{xy}$ as a function of the number of writing current pulses in Co/Pt bilayers with $H = 0$. The measurement configuration is identical to that of YIG/Pt sample. The polarity of $\Delta R_{xy}$ variation of Co/Pt is opposite to that of YIG/Pt. (d) Schematic of current-induced magnetization switching in YIG/Pt, where $M_1$ and $M_2$ denote magnetic moments in two sublattices. $M_1$ and $M_2$ are switched toward the writing current direction.

We then turn toward the current-induced switching mechanism of YIG/Pt. Note that only one ferrimagnetic insulator layer YIG is used, hence SOT rather than spin-transfer torque exists in our case. Remarkably, the switching polarity of $\Delta R_{xy}$ in Fig. 1(e), negative for write 1 and positive for write 2, reveals that the sublattice magnetizations in YIG are aligned parallel/antiparallel to the direction of writing current according to SMR theory [41]. This feature is also supported by the longitudinal resistance variation [40]. These experimental observations unravel that the current-induced in-plane sublattice magnetizations switching of bi-axial ferrimagnet YIG/Pt is in analogy to the Néel order switching toward the writing current direction in antiferromagnet/Pt systems [35,37], where the antidamping-torque dominates. It is also reasonable to propose that the antidamping-torque may induce the ferrimagnetic moments switching of bi-axial YIG, where two sublattices with antiparallel magnetic moments is strongly antiferromagnetic coupled, though uncompensated. Note that the polarity of current-induced switching in ferrimagnetic amorphous metal CoGd [19], which possesses negligible magnetic anisotropy, is opposite to our results in bi-axial YIG. This indicates that the magnetic anisotropy in ferrimagnet has a vital effect on current-induced in-plane switching, which is supported by the absence of SOT-induced switching in our YIG(111) samples [40].

To further study the SOT switching in biaxial ferrimagnetic YIG where the magnetization is aligned parallel/antiparallel to the current axis, we performed control experiment with ferromagnet and compared the current-induced switching polarity of $\Delta R_{xy}$ using Co (2 nm)/Pt (5 nm) bilayers where the magnetization should be aligned perpendicular to the current axis [17,18,45]. Although Co is a conductor, which is not perfect as a comparison of YIG, Co/Pt bilayer possesses considerable SMR signal [46], from which the direction of magnetization can be readout easily. Therefore, Co/Pt bilayer used as a control sample is reasonable. The measurement configuration is identical to that of YIG/Pt [Fig. 1(c)]. The variation of $\Delta R_{xy}$ in Fig. 2(c) is opposite to the YIG/Pt case. Once an external field of 10 kOe is applied on the devices, the switching signals vanishes [40], indicating the variation of $\Delta R_{xy}$ is indeed due to the switching of Co moments. In the case of current-induced in-plane switching of ferromagnet, the magnetic moment should be switched to the direction of spin polarization [17,18,45], which is perpendicular to the writing current direction. This process can be understood by the transfer of spin angular moment from the spin-polarized current to the magnetization, similar to the spin-transfer-torque scenario. This control experiment further confirms that the sublattice magnetizations of YIG are aligned parallel/antiparallel to the writing current direction, rather than being aligned along spin polarization direction. This may indicate the importance of antiferromagnetic coupled sublattices during current-induced in-plane switching in bi-axial ferrimagnet. The charge current in the Pt layer produces spin current and the resultant spin accumulation by spin Hall effect [11], exerting a torque on the two anti-parallel magnetic sublattices ($M_1$ and $M_2$) and then resulting in their switching parallel or antiparallel to the writing current direction [Fig. 2(d)]. The intriguing in-plane switching feature in our case discloses that as a typical FiMI, the anti-paralleled magnetic sublattice in YIG may play an important role, which is similar to the Néel order switching in antiferromagnets to some extent at least from the current-induced in-plane magnetization switching viewpoint. More experiments or simulations in different FiMI are needed for further understanding of current-induced in-plane switching in FiMI systems. Based on those results and analyses, the magnetic field manipulated true current-induced switching and co-existence of artifacts [40] make in-plane bi-axial ferrimagnet YIG a model for investigating the current-induced switching, which could promote the application of ferrimangetic spintronics.

In addition to the SOT-induced magnetization switching, we have explored SOT-induced magnetic moments tilting toward the current direction by SMR measurements with different $J$ and $H$. For these experiments, Hall resistance ($R_{xy}$) was recorded when the current was applied along one of the easy-axes [110] and the magnetic field was in-plane rotated from the current ($I$) direction (angle $\alpha_H = 0$ for $H \parallel I$), as depicted in Fig. 3(a). Figure 3(b) shows the angle $\alpha_H$ dependent $R_{xy}$ with different current densities ($J = 1.2, 2, 6, 8,$ and $10 \times 10^6$ A cm$^{-2}$) and $H = 50$ Oe. Remarkably, as the current density increases gradually from 1.2 to $10 \times 10^6$ A cm$^{-2}$, within 0–90 ° scale the high Hall resistance state (peak) shifts to a high rotation angle. On the basis of the SOT-switching results described above, magnetization is tilted toward the current direction, therefore $M_1$ ($\alpha_M$) and $M_2$ deviate from the field direction and toward the current direction ($\alpha_M = 0$) [Fig. 3(c)]. Thus it is necessary to use a magnetic field with a larger rotation angle $\alpha_H$ to compensate the tilting tendency induced by SOT, which results in the shift of SMR curves. Also visible is that in the range of 90–180 ° the low Hall resistance state (valley) shifts to a low rotation angle, because the SOT induces magnetization switching toward the current direction ($\alpha_M = 180$ °) and then a magnetic field with a smaller rotation angle $\alpha_H$ is employed. The variation tendencies in 180–270 ° (peak) and 270–360 ° (valley) scales are similar to these two scenarios, respectively. In general, the SMR curve exhibits a high and low Hall resistance states for $\alpha_H = 45$ °/225 ° and 135 °/315 °, respectively [41]. Note that the slight deviation of the observed peak and valley with low current density [Fig. 3(b)] from these theoretical angles is due to a low field of 50 Oe used, which is below the saturation field of the YIG film. Such a deviation vanishes when a high magnetic field is used, such as $H = 5000$ Oe [40].

The angle shift of the SMR curves as a function of $J$ with different external fields ($H = 50, 100, 1000,$ and 5000 Oe) is summarized in Fig. 3(d), where $\Delta\alpha_H$ is the angle difference between the valley and peak (in 0–180 °) of $R_{xy}$. There are two striking features in the figure: (i) The angle difference $\Delta\alpha_H$ is reduced (angle shift is enhanced) with increasing $J$; (ii) The modulation of $\Delta\alpha_H$ is suppressed with increasing magnetic field, which almost vanishes with a high *H*, e.g., 5000 Oe [40]. It is easy to understand that a larger current induces stronger magnetization titling toward the current direction, resulting in a larger "compensated angle" for the field, which reduces $\Delta\alpha_H$. With increasing magnetic field, the SOT-induced magnetization tilting is negligible because the magnetization is always retained along the field. The magnetization switching phenomena deduced from these SMR results with different *J* and *H* are consistent with switching responses in Fig. 2. Thereby the current-dependent SMR measurements support SOT-induced in-plane sublattice magnetizations switching toward parallel/antiparallel to the writing current direction. In addition, we quantify the SOT-field equivalence through the summary of $\Delta\alpha_H$ resulting from SOT induced SMR tilting. An extra angle $\alpha_{SOT,J}$ produced by SOT with current density *J* is introduced in transverse SMR equation:

$$R_{xy} = \Delta R \sin 2(\alpha_H - \alpha_{SOT,J}), \ 0° < \alpha < 90° \quad (1)$$

$$R_{xy} = \Delta R \sin 2(\alpha_H + \alpha_{SOT,J}), \ 90° < \alpha < 180° \quad (2)$$

where $R_{xy}$ and $\Delta R$ are transverse resistance and amplitude of transverse SMR, respectively. When $\alpha_H = 45°+\alpha_{SOT,J}$ ($135°-\alpha_{SOT,J}$), $R_{xy}$ is high (low) resistance state, namely peak (valley) of SMR in 0–180°. Therefore, $\Delta\alpha_H$ in Fig. 3(d) equals $90°-2\alpha_{SOT,J}$. The variation of $\Delta\alpha_H$ with different *J* under *H* = 50 Oe are fitted by linear functions [dashed line in Fig. 3(d)], and the slope *k* is around $-4.1$ deg/($10^6$ A cm$^{-2}$). Based on the equation of slope *k*:

$$k = \frac{(90°-2\alpha_{SOT,J1})-(90°-2\alpha_{SOT,J2})}{J1-J2} = \frac{-2\alpha_{SOT,J1}+2\alpha_{SOT,J2}}{J1-J2} = \frac{-2\Delta\alpha_{SOT,J}}{\Delta J} \quad (3)$$

where $\alpha_{SOT,J1(J2)}$, $\Delta\alpha_{SOT,J}$ and $\Delta J$ are angles induced by SOT with current density *J*1(*J*2), the variation of $\alpha_{SOT,J}$ and the change of current density *J*, respectively, the value of $\frac{\Delta\alpha_{SOT,J}}{\Delta J}$ is calculated to be around 2.1 deg/($10^6$ A cm$^{-2}$). Based on the above analyses, the sublattice magnetization is switched toward the direction of current, hence the SOT induced equivalent field $H_{SOT}$ is set along current axis here [40]. On the basis of magnetic field vector addition, the SOT-field equivalence in YIG is determined to be 2.6 Oe/($10^6$ A cm$^{-2}$) [40], which is comparable with or slightly larger than the value obtained in typical ferrimagnetic insulators, such as TmIG [0.6 Oe/($10^6$ A cm$^{-2}$)] [26].

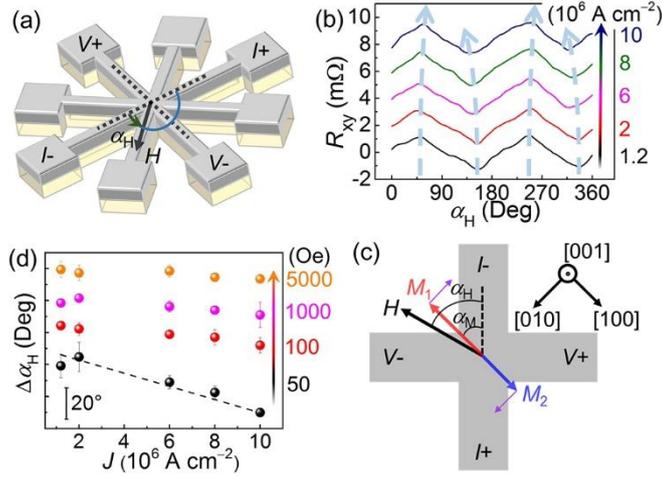

FIG. 3. Current-induced magnetization tilting during SMR measurements. (a) Measurement configurations of SMR. The current $I$ is applied along [110]. The magnetic field $H$ rotates in the plane of device. (b) SMR curves with different current densities (marked in the inset) and $H$ = 50 Oe. The dashed arrows are a guide to reflect the shift of peak/valley positions. (c) Schematic of the magnetization tilting of two sublattices of YIG induced by applied current. $\alpha_M$ ($\alpha_H$) denotes the angle between $I$ and $M_1$ ($H$). The red ($M_1$) and blue ($M_2$) arrows denote the magnetization of two magnetic sublattices, and the thin purple arrow represents the tilting direction. (d) Summary of the angle difference $\Delta\alpha_H$ between the valley and peak (in 0–180°) of SMR with different $J$ and $H$. The typical $H$ used are marked. The error bars are estimated from standard deviation of three SMR measurements. The dashed line is linear fitting of $\Delta\alpha_H$ with different current densities under $H$ = 50 Oe.

We then focus on the YIG-thickness ($t$) dependent transport measurements. Figure 4(a) presents the SOT-induced $\Delta R_{xy}$ variation for 15, 20, 30, and 60 nm-thick YIG. All of the YIG/Pt samples exhibit reversible $\Delta R_{xy}$ variation. A comparison of the $\Delta R_{xy}$ shows that the magnitude of the $\Delta R_{xy}$ is greatly enhanced with increasing $t$ to 30 nm, and then $\Delta R_{xy}$ is saturated and keeps almost unchanged even up to 60 nm. On the other side, the angle $\alpha_H$ dependent SMR curves measured with $H$ = 5000 Oe for different $t$ is shown in Fig. 4(b). Remarkably, the magnitude of SMR signals are enhanced with increasing $t$ and saturated at $t$ = 30 nm, which coincides with the thickness-dependence of SOT-induced $\Delta R_{xy}$ variation. Since SOT-induced switching results in the present case are dependent on two factors, the current-based writing efficiency and the SMR-based readout capability, it is significant to exclude the influence of thickness-dependent SMR when exploring the switching efficiency. In this scenario, the ratio of $\Delta R_{xy}/\Delta$SMR, where $\Delta$SMR is the $R_{xy}$ difference between the peak and valley of SMR curves in Fig. 4(b), is introduced to reflect the switching efficiency in our case. The ratio of $\Delta R_{xy}/\Delta$SMR as a function of $t$ is displayed in Fig. 4(c), which shows a gradual enhancement with increasing $t$ and is almost saturated at $t$ = 30 nm. Meanwhile, the saturation magnetization ($M_S$) [40] of the YIG films is also presented in Fig. 4(c) for a comparison. It is found that both of them show a similar thickness-dependence, suggesting that the switching efficiency is relevant to $M_S$. Because of the enhancement of $M_S$ and corresponding interfacial exchange interaction, more spin current can flow into the YIG [28], which enhances switching efficiency of YIG and resultant $\Delta R_{xy}/\Delta$SMR.

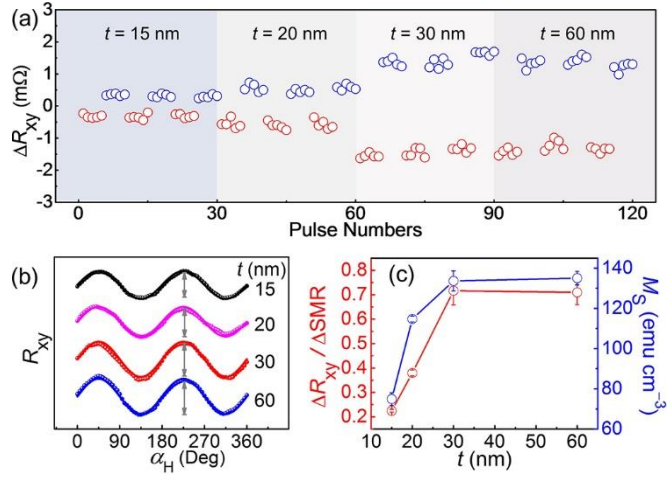

FIG. 4. SOT-induced switching and SMR measurements in YIG/Pt with different YIG thicknesses ($t$). (a) $\Delta R_{xy}$ as a function of pulse numbers in YIG/Pt bilayers with $t$ = 15, 20, 30, and 60 nm. Results of different $t$ are separated by regions of different colors. (b) SMR results with different $t$ under $H$ = 5000 Oe. The $R_{xy}$ differences between peak and valley ($\Delta$SMR) are denoted by (grey) arrows in the inset, and the values are 3.0, 3.2, 3.8, and 3.8 m$\Omega$ for $t$ = 15, 20, 30, and 60 nm, respectively. (c) $\Delta R_{xy}/\Delta$SMR and $M_S$ versus $t$. Both $\Delta R_{xy}/\Delta$SMR and $M_S$ show a similar thickness-dependence and saturate at $t$ = 30 nm. The error bars of $\Delta R_{xy}/\Delta$SMR and $M_S$ are estimated from standard deviation of reversible switching and three magnetization measurements, respectively.

## IV. CONCLUSION

In summary, we have demonstrated the reversible in-plane magnetization switching of YIG in YIG/Pt bilayers by SOT. The switching signal is readout by spin Hall magnetoresistance. The sublattice magnetizations of YIG are found to be aligned parallel/antiparallel to the direction of writing current, and may be ascribed to the antidamping-torque for the two strongly antiferromagnetic coupled sublattices, which is similar to the Néel order switching in antiferromagnetic system to some extent. This phenomenon indicates that anti-paralleled sublattices in ferrimagnetic insulator may play an important role during the current-induced in-plane magnetization switching process, and more studies with different materials are needed to further explore switching in ferrimagnetic insulators in the future. Our finding not only promotes the understanding of current-induced switching, but also accelerates combination of current-induced magnetization switching and previous microwave devices/spin caloritronics to realize high energy efficient spintronic applications based on magnetic insulators modulated by all-electrical means.

**ACKNOWLEDGMENTS**

We thank Dr. D. Z. Hou for helpful discussions. C.S. acknowledges the support of the Beijing Innovation Center for Future Chips, Tsinghua University and the Young Chang Jiang Scholars Programme. This work was supported by the National Key R&D Programme of China (grant nos. 2017YFB0405704 and 2017YFA0206200) and the National Natural Science Foundation of China (grant nos. 51871130, 51571128, 51671110 and 5183000528).

YIG(001)/Pt, for SOT-field equivalence deduced from SMR tilting and for hysteresis loops of YIG(001) with different thicknesses, which includes Ref. [41].